\documentclass[prb,twocolumn,aps,showpacs,letter]{revtex4}

\usepackage{graphicx}
\usepackage{bm}
\usepackage{amssymb} 
\usepackage{amsmath}
\usepackage{subfigure}
\usepackage{epstopdf}
\usepackage{bigints}

\begin{document}


\title{Synthesis of thin silicon dioxide layers with high E' center densities and investigation of the E' center spin relaxation dynamics for single spin readout applications}
\author{K. Ambal}
\author{A. Payne}
\author{D. P. Waters}
\author{C.C. Williams}
\email[]{clayton@physics.utah.edu}
\author{C. Boehme}
\email[]{boehme@physics.utah.edu}
\affiliation{Department of Physics and Astronomy, University of Utah, Salt Lake City, UT 84112}


\begin{abstract}
   Methods for the creation of thin amorphous silicon dioxide (aSiO$_2$) layers on crystalline silicon substrates with very high densities of silicon dangling bonds (so called E' centers) have been explored and volume densities of [E']$>5\times 10^{18}\mathrm{cm}^{-3}$ throughout a 60nm thick film have been demonstrated by exposure of a thermal oxide layer to a low pressure Argon radio frequency plasma. While the generated high E' center densities can be annealed completely at $300^o$C, they are comparatively stable at room temperature with a half life of about one month. Spin relaxation time measurements of these states between $T=5$K and $T=70$K show that the phase relaxation time $T_2$ does not strongly depend on temperature and compared to SiO$_2$ films of lower E' density, is significantly shortened. The longitudinal relaxation time $T_1\approx 195(5)\mu\mathrm{s}$ at room temperature  is in agreement with low--density SiO$_2$. In contrast, $T_1\approx 625(51)\mu\mathrm{s}$ at $T=5$K is much shorter than in films of lower E' density. These results are discussed in the context of E' centers being used as probe spins for spin--selection rules based single spin--readout.
\end{abstract}




\maketitle


\section{Introduction}
 In thermal equilibrium, amorphous thin film silicon dioxide (a-SiO$_2$) can contain large quantities of highly localized silicon dangling bond states, so called E' centers~\cite{lenahan1998can,lenahan2000centers}. These defects exhibit positive correlation energies and are therefore paramagnetic, a property that allows us to study these centers with electron paramagnetic resonance (EPR) spectroscopy techniques~\cite{Atherton1993,Schweiger2001}. As E' centers limit the performance of a-SiO$_2$ device components (e.g. the gate dielectric of silicon thin film transistors~\cite{lenahan2000centers}), most studies of E' center properties have focused primarily on how E' center densities can be minimized by a-SiO$_2$ preparation and treatment. Few studies in the past have focused on the dynamic properties of this spin $s=1/2$ system, but those that have show that E' centers exhibit remarkably long longitudinal ($T_1$) spin relaxation times over large temperature ranges~\cite{PhysRev.130.577,castle1965temperature,eaton1993irradiated,ghim1995magnetic}. At room temperature $T_1$ times on the order of hundreds of microseconds have been reported~\cite{eaton1993irradiated}. This is long compared to the $T_1$ times of silicon dangling bonds at the a-SiO$_2$ to crystalline silicon (c-Si) interface~\cite{PhysRevB.81.075214} (the so called P$_\mathrm{b}$ centers)and it is comparable to other bulk silicon dangling bonds in amorphous silicon~\cite{PhysRevB.28.6256} or microcrystalline silicon~\cite{malten1997pulsed,boehme2003dynamics}. Hence, even though E' centers are generally viewed as detrimental for technological applications, they could become important for spin applications such as spin memory concepts or as probe spins for spin readout applications~\cite{Boehme2002}. In fact, recently, it was shown that E' centers close to the c-Si/a-SiO$_2$ interface can be used as probe spins for the phosphorous donor nuclear/electron spin qubit~\cite{paik2011electrically}, an intensively studied impurity qubit system which is currently among the most coherent known quantum systems in nature~\cite{tyryshkin2011electron}.

 Spin pair based qubit readout requires that the distance between the probe and the test spin can be shifted relative to its respective pair partner on atomic or possibly subatomic length scales~\cite{Boehme2002}. While experimental demonstrations have been given that this qubit--probe spin pair readout works extraordinary well when the two states have an appropriate proximity, it has remained elusive so far how to establish and manipulate such distances in a controlled and reproducible manner. We are currently in the pursuit of achieving such distance control by means of scanning probe microscopy, an idea that requires scanning probes with an a-SiO$_2$ surface and a single E' center at its tip. Since E' centers develop at random sites within the continuous random network of a-SiO$_2$, we suggest to fabricate individual probe spins at cantilever tips by growth of an a-SiO$_2$ layer on c-Si cantilevers whose E' density is large enough such that a sufficiently large probability exists that a single E' center is very near the apex of the tip.

 The silicon dioxide needed for the spin readout measurements must meet four criteria: (1) The density of E' centers must be high enough such that the probability to find a center at the apex of a cantilever probe tip is of the order of unity. An estimate for the active tip volume in which a single E' center can be utilized as a probe spin in the grown oxide layer can be obtained from the product of the tip surface area of less than 300nm$^2$ for a 25nm tip radius and a tunneling depth of less than 2nm. Thus, the oxide layer requires E' densities between 10$^{18}$cm$^{-3}$ and 10$^{19}$cm$^{-3}$. This is higher than the highest previously reported E' densities~\cite{lenahan2000centers} which were generated via electric currents through silicon dioxide gate dielectrics, a procedure that is hardly applicable to cantilever surfaces. (2) The E' centers at high densities must exhibit similarly long transverse spin relaxation times ($T_1$) as at low densities. If the proximity of the E' states significantly increases the spin relaxation, applicability for spin readout will be limited~\cite{Boehme2002}. (3) The E' centers must be stable at room temperature and under ambient light illumination. They need to be chemically stable. Either limited defect lifetimes or limited spin lifetimes (relaxation times) could equally be showstoppers for the applicability of E' centers to single spin readout schemes. (4) The electrons from the E' centers should not leak from one center to another.

 In the following sections, we report unsuccessful and successful attempts to create SiO$_2$ layers with very high densities of E' centers. The thermal and light exposure stability of the high E' densities will be discussed and the relaxation dynamics of E' spins in very high density layers will be presented (for both, $T_1$ and $T_2$ times). The results of these studies will then be discussed with regard to the suitability of this newly developed high E' density SiO$_2$ for scanning probe controlled single spin detection and readout.


\section{Synthesis of thin $\mathbf{a-SiO_2}$ films with very high E' center densities}
\begin{figure}
\includegraphics[width=8.5cm]{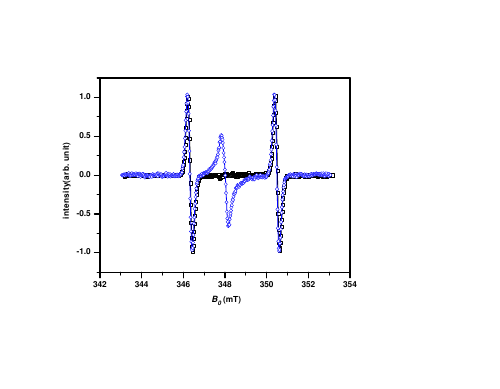}
\caption{Plots of X-Band EPR spectra of 60mm x 3mm x 0.3mm large $^{31}$P doped c-Si(111) samples measured at a temperature $T=20$K with a field modulation frequency $f=10$kHz, a modulation amplitude of 0.1mT, and a weak microwave power of 4$\mu$W to avoid saturation. The samples had 60nm thin layers of thermally grown a-SiO$_2$. The black data points represent measurements of the as prepared thermal oxide. The blue data points show a measurements under identical conditions after the sample was been exposed to an argon ion plasma for 5 minutes.}
\label{fig1}
\end{figure}

For the study of various E' preparation techniques, we used n-type, phosphorous doped ([$^{31}$P]$\approx 10^{15}$cm$^{-3}$) Czochralski grown c-Si(111) wafers. The use of phosphorous doped material allowed a very accurate determination of the E' densities from EPR spectra since the well known hyperfine split $^{31}$P resonance could be used as an in-sample spin-standard. The 300 micrometer thick 3" wafers were first annealed in oxygen at atmospheric pressure and 1000$\mathrm{^oC}$ in order to form an approximately 60nm thick (profilometer measured) thermally grown a-SiO$_2$ layer. The oxidized samples were then diced into 60mm x 3mm size EPR compatible rectangles. The black data points of Fig.~\ref{fig1} display an EPR spectrum of the as prepared thermal oxides. The two peaks are due to the hyperfine split $^{31}$P resonance that, due to the known phosphorous bulk density and therefore areal density ($\approx3\times10^{13}\mathrm{cm}^{-2}$), can be used to calibrate the magnetic field and density scales for the E' center measurements. The black data points show almost no resonant features next to the phosphorous hyperfine peaks which means that E' densities in as grown samples are below the detection limit which is $\approx 1\times 10^{12}\mathrm{cm}^{-2}$ for the given measurement conditions. This corresponds to an average volume density below $\approx1.67\times10^{17}\mathrm{cm}^{-3}$ within the 60nm thin film. Thus, given that previous reports of E' densities in thermally grown SiO$_2$ are all significantly below the $10^{17}\mathrm{cm}^{-3}$ limit, we expected no significantly different EPR signals from E' centers for the as grown oxide layers.

In order to explore how to create E' densities $>10^{17}\mathrm{cm}^{-3}$, the thin a-SiO$_2$ layer was exposed to (i) ultraviolet (UV) radiation~\cite{yokogawa1990positive} (produced by a NdYAG laser with 264nm wave length) for six hours, (ii) gamma radiation produced by a $^{137}$Cs sample for 24 hours, producing an overall irradiation dose of about 10-12Mrad~\cite{Zvanut1992}, (iii) different growth temperatures during the thermal growth, (iv) an Ar-ion discharge plasma excited by a 300W 13.56MHz RF excitation at 0.5sccm gas flow and a pressure of 10mTorr~\cite{ishikawa2005reduction,ichihashi2006effects}. We then conducted EPR measurements similar to those shown in Fig.~\ref{fig1} on the samples treated according to (i), (ii), and (iii). These measurements revealed similar results compared to the as grown sample, represented by the black data in Fig.~\ref{fig1}. This again confirmed the previous reports that treatment of a-SiO$_2$ layers following these methods may increase the E' center densities but not beyond the $10^{17}\mathrm{cm}^{-3}$ range.

\begin{figure}
\includegraphics[width=8.5cm]{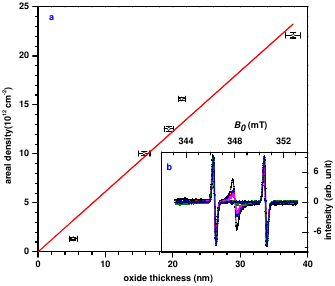}
\caption{(a) Plot of the measured E' center area density as a function of the different oxide thicknesses and an offset--free linear fit (red line). The agreement of the measured data and the linear fit indicates that the observed paramagnetic defects created by the Ar$^+$ plasma are bulk defects. (b) Plots of the EPR spectra measured on a-SiO$_2$ samples that have been exposed for different durations to dilute HF. The remaining oxide thickness on each sample was measured by ellipsometry. }
\label{fig2}
\end{figure}

In contrast, the application of the Ar-ion plasma treatment (method iv) caused a significant increase of the E' density, as indicated by the blue circled data points in Fig.~\ref{fig1}. The plot displays a feature at a magnetic field of approximately 348mT, corresponding to a Land\'e-factor of $g\approx 2.001$ which is attributed to plasma induced E' centers. The average E' center volume density in this film derived from the measured areal density per film thickness is $6\times 10^{18}$cm$^{-3}$, determined by using the phosphorus donor spins in the silicon substrate as a  reference. While this observation shows that the plasma exposure of the oxide film is able to generate a large quantity of paramagnetic species at the $g$-factor anticipated for E' centers, it is not clear whether these states are all E' centers (silicon dangling bonds within the SiO$_2$ bulk). Other paramagnetic species such as interface defects between the SiO$_2$ layer and the c-Si bulk (if the etch has not removed the entire oxide) or the plasma etched c-Si surface states (if the oxide was completely removed) could also contribute to the observed signal. In order to explore this question, the depth distribution of the plasma induced paramagnetic states was studied by repetition of the EPR determined density measurement as a function of several oxide thicknesses after the partial removal of the oxide by a wet chemical etch. For this step etch experiment, a dilute HF solution was used. After each HF-etch step, the oxide thickness was measured by ellipsometry and the areal density of the paramagnetic centers was determined by EPR spectroscopy. Figure~\ref{fig2} displays the results of these measurements for both the area concentration as a function of remaining oxide thickness (a) and the raw data given by the EPR spectra of the sample recorded after the individual etch steps (b).

Plot (a) also displays an offset--free linear fit which shows good agreement with the data. This agreement is indicative of a homogeneous distribution of the plasma--generated centers throughout the oxide layer. From the slope of the fit, we obtain a volume density $6.2(3)\times 10^{18}\mathrm{cm}^-3$. Based on the measurements presented in Fig.~\ref{fig2}, we conclude that we have found a method to generate SiO$_2$ layers with very large densities of paramagnetic E' centers as needed.


\section{Thermal and light induced stability of very high E' center density film}
\begin{figure}
\includegraphics[width=8.5cm]{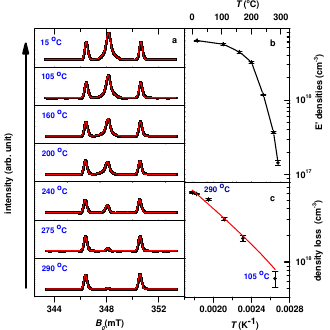}
\caption{(a) EPR spectra measured after different annealing temperatures (black lines) and fit of the data with three Gaussian peaks. The density of the E' centers decreases with temperature and at $290^o$C, it is reduced by an order of magnitude. (b) Plot of the E' center densities obtained from the fit results displayed in (a) as a function of the temperature. (c) Arrhenius plot of the density loss, the difference of the room temperature sample and the annealed samples as a function of the anneal temperature. The fit with an Arrhenius function reveals a resonable agreement and a defect anneal activation energy of 0.176(1)eV.}
\label{fig3}
\end{figure}

In order to study the thermal stability of the large Ar$^+$ plasma induced E' center densities, we conducted a series of anneal experiments on high density samples that were plasma treated for 5 minutes with the plasma parameters described above. The thermal anneal was then conducted for 20min under ambient conditions at various temperatures between room temperature and 290$^o$C. Using EPR, the E' center's area density was then measured as described above. The results of these measurements are displayed in Fig.~\ref{fig3}(a). The set of spectra illustrates how the plasma generated ensemble of paramagnetic states gradually disappears with increasing anneal temperature. The plot in Fig.~\ref{fig3}(b) displays the E' center densities that were derived from the EPR measurements as a function of the preceding anneal temperatures. From the difference of the E' densities of the non-annealed sample and this data, one can obtain the density loss as a function of temperature, which is displayed as an Arrhenius plot in Fig.~\ref{fig3}(c). The fit of this data with an Arrhenius function reveal an activation energy of $\Delta=176(1)$meV. The anneal experiments show that plasma induced high E' center densities can be annealed at comparatively low temperature. However, since $\Delta>k_\mathrm{B}T_\mathrm{room}$, room temperature stability of the defects is observed.

\begin{figure}
\includegraphics[width=8.5cm]{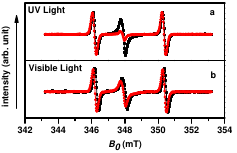}
\caption{EPR spectra of high E' center density SiO$_2$ films measured before (black data) and after (red data) a one hour exposure with UV light (a) and visible light (b). Both photo--bleaching experiments show that the light exposure leads to a reduction of the E' center densities. However, this effect is significantly stronger for UV light exposure.}
\label{fig4}
\end{figure}

In order to further scrutinize the stability of the plasma induced high E' center densities, we have conducted photo--bleaching experiments. We exposed plasma treated but not annealed SiO$_2$ layers for 60 minutes to two different light sources: (a) a UV source with two strong emission maxima at around 174nm and 254nm, and (b) an incandescent spectral light lamp which emits mostly in the visible wavelength range. Fig.~\ref{fig4} displays two plots, each of which contain two EPR spectra of the plasma etched but otherwise untreated sample and the bleached samples, respectively. The two plots (a) and (b) correspond to the UV bleaching experiment and the visible light experiment, respectively. The data sets show that photo bleaching has a significant effect for both light sources as both post exposure spectra exhibit smaller E' center resonances. However, in comparison to the comparatively minor loss for the visible spectral lamp (b), the exposure by UV light causes a reduction of the E' density by a significantly larger amount. This realization that bleaching can have similar effects as annealing could be significant for the development of low--temperature adjustment of well defined E' center densities.

\begin{figure}[b]
\includegraphics[width=8.5cm]{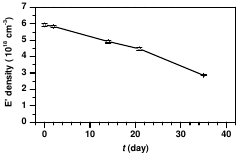}
\caption{Plot of EPR measured E' center densities of an a-SiO$_2$ layer after plasma treatment over the course of approximately five weeks. A gradual decline of the density is observed. However, the decay is slow enough such that even after about five weeks the absolute volume density still exceeds $10^{18}\mathrm{cm}^{-3}$.}
\label{fig5}
\end{figure}

Finally, we tested the long term stability of the plasma generated high E' center densities at room temperature. Using EPR, we measured repeatedly the density of a plasma treated sample over a time of approximately five weeks. During this time, the sample was kept at ambient conditions and at room temperature. The results of these measurements are displayed in the plot of Fig.~\ref{fig5}. Over the course of about month, a clear decline of the E' density to about half of its original value is recognizable. While this is a significant decrease, the resulting half life of the generated E' center densities exceeds by far the expected duration of the single spin experiments for which the high E' center densities are needed.

\section{Spin relaxation times of E' centers at high densities}
The application of high density E' center SiO$_2$ layers for scanning probe based spin readout requires sufficiently long spin relaxation times $T_1$ and $T_2$. The magnetic resonance spectra discussed above give no indication that there is any microscopic difference between E' centers in the high density material reported here compared to previously studied low density materials. Consequently, one may hypothesize that the intrinsic relaxation behavior of an individual E' center could be similar or identical in high- and low-density films. However, the decreased average density between the E' centers at high densities could increase their mutual spin interactions, mostly because of spin--dipolar coupling, not because of exchange since the latter is weak due to the strong localization~\cite{lenahan1998can} of the E' center. Since spin--spin interaction can quench relaxation times, an experimental study of the high density E' center relaxation times is necessary.

\begin{figure}
\includegraphics[width=8.5cm]{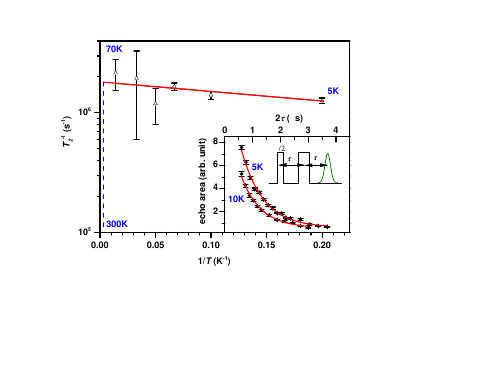}
\caption{Plot of the measured transverse E' center spin relaxation rate coefficients $T_2^{-1}$ and their error margins as a function of the inverse temperature. The red line represents a fit with an Arrhenius function. The inset displays a sketch of the Hahn--spin echo sequence that was used to measure $T_2$ as well as a plot of the measured Hahn--echo intensity as a function of the pulse separation time $\tau$ for $T=5$K and $T=10$K, with the plots of fits of these data sets with exponential decay functions. Within the given error margins, no temperature dependence of the $T_2$ relaxation is observed.}
\label{fig6}
\end{figure}

\begin{figure}
\includegraphics[width=8.5cm]{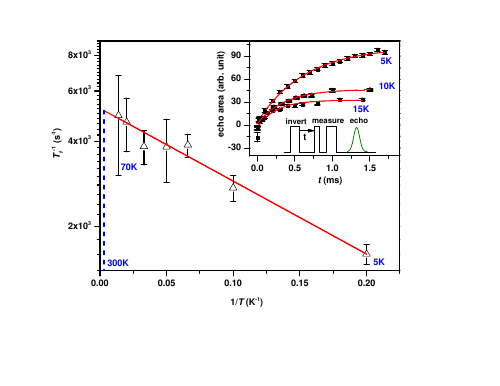}
\caption{Plot of the measured longitudinal E' center spin relaxation rate coefficients $T_1^{-1}$ and their error margins as a function of the inverse temperature. The red line represented a fit with an Arrhenius function. The dashed line indicates the room temperature value of $T_1$  that is extrapolated from the measurements conducted at lower temperatures. The inset displays a sketch of the inversion recovery pulse sequence that was used to measure $T_1$ as well as a plot of the Hahn-echo intensity as a function of the inversion delay time $t$ for $T=5$K, $T=10$K, and $T=15$K as well as plots of the fit results with exponential recovery functions.}
\label{fig7}
\end{figure}

Figures.~\ref{fig6} and \ref{fig7} show the results of both longitudinal ($T_1$) and transverse ($T_2$) spin relaxation times on the high density SiO$_2$ reported above. For these measurements, we have applied pulsed EPR experiments in a temperature range of $T=5$K to $T=70$K. Due to the temperature dependence of equilibrium polarization, spin echo measurements could be conducted on the very small spin ensemble of the thin film samples only up to about 70K. However, as shown in Figs.~\ref{fig6} and \ref{fig7}, the obtained data shows good agreement with Arrhenius functions, and therefore, an extrapolation of the experimental data to room temperature appeared viable and is indicated in the figures.

In order to measure $T_2$ relaxation times, a two pulse Hahn--echo experiment was performed. Figure~\ref{fig6} displays the results of these measurements (the relaxation rate coefficient $T_2^{-1}$) as a function of the inverse temperature ($1/T$). The data points in this plot were obtained by execution of Hahn--echo decay experiments where a standard Hahn-echo pulse sequence consisting of a $\pi/2-\pi$ is applied on resonance to the spin ensemble and the integrated intensity of the resulting spin--echo is then measured as a function of the pulse separation time $\tau$. For the examples at low temperature ($T=5$K, $T=10$K), the employed pulse sequence as well as the decay data of the Hahn--echo amplitude are displayed in the inset, along with a fit by an exponential decay function which shows excellent agreement with the experimental data. The resulting decay constants for these temperatures as well as the other temperatures displayed in the main plot were fit with an Arrhenius function.

The measured transverse spin relaxation times of $T_2\approx 0.5\mu$s showed no significant dependence on temperature. These measurements display significantly shorter $T_2$ times compared to room temperature values obtained on bulk SiO$_2$ ~\cite{ghim1995magnetic,Motoji1992}. This suggests that the significantly higher densities of the thin film materials investigated here cause an increase in spin-spin interaction between E' centers, and thus, the transverse spin-relaxation times are shortened. In contrast, the temperature independence is indicative that phonon-processes do not play a role for $T_2$.

For the measurements of the longitudinal spin relaxation times $T_1$, the Hahn-echo pulse  sequence used for the $T_2$ measurements was extended by one pulse such that polarization inversion recovery could be observed. The inset of Fig.~\ref{fig7} displays this pulse scheme which begins with a $\pi$-inversion pulse of the spin ensembles equilibrium polarization. After the inversion, a delay time $t$ passes before a Hahn--echo pulse sequence is applied which reveals the residual polarization of the spin ensemble. Measurement of the ensemble polarization as a function of the delay time $t$ will then reveal the dynamics of how the inverted spin polarization right after the inversion pulse gradually relaxes back towards a thermal equilibrium polarization. The inset of Fig.~\ref{fig7} shows plots of the measured polarization as a function of the delay time $t$ for temperatures $T=5$K, $T=10$K, and $T=15$K. The data sets show that while for small $t$, the measured residual polarization is less than 0, representing a negative (inverted) polarization, a positive equilibrium polarization is reestablished for long $t$. This experimental data is well fit by exponential recovery functions and the time constants obtained from these fits represent the measured $T_1$ values. The main plot of Fig.~\ref{fig7} displays the measured $T_1$ rate coefficients ($T_1^{-1}$) as a function of the inverse temperature as well as the plot for an Arrhenius function that has been fit to the experimental data. The extension of this Arrhenius function to $T=300$K reveals an extrapolated room temperature longitudinal spin relaxation time of $T_1=195(5)\mu$s, a value that is in good agreement with $T_1$ times measured on low density bulk SiO$_2$~\cite{eaton1993irradiated,ghim1995magnetic}. At low temperatures, the longitudinal spin relaxation times vary significantly from measurements made on low E' center--density SiO$_2$~\cite{castle1965temperature}. We hypothesize that the density dependence of the $T_1$ times at low temperatures is caused by dominating spin--spin interactions which increases with decreased average E' center distances at higher densities. In contrast, at high temperatures when phonon densities are high, the longitudinal relaxation appears to be dominated by spin--lattice processes, not spin--spin interactions. Therefore, $T_1$ will exhibit no dependence on the E' center density under these conditions.


\section{Summary and Conclusions}
Very high densities ($n[E']>5\times10^{18}\mathrm{cm}^{-3}$) of paramagnetic E' centers in thin silicon dioxide films have been generated using Ar+ ion plasma etching. The defects exhibit a homogeneous density within 60nm of the surface, and at room temperature, they are stable on time scales of about one month. The high--density E' centers can be removed by annealing with a decay activation energy of $\Delta=176(1)$meV and by light exposure with wavelengths below 254 nm. Measurements of both the $T_1$ and $T_2$ spin relaxation times of E' centers showed similarities of the high density E' centers compared to low densities. Between $T=5$K and $T=70$K, the spin-spin relaxation time ($T_2$) is ~552(15)ns, independent of the materials temperature. The spin-lattice relaxation time ($T_1$) exhibits a temperature dependence within the same temperature range and by extrapolation through an Arrhenius function, a room temperature $T_1\approx195(5)\mu$s is estimated. This value is is good agreement with literature values for low E' density materials. At $T=5$K, a $T_1=625(51)\mu$s was observed for the high--density SiO$_2$ which is significantly shorter than the low--density $T_1$\cite{castle1965temperature}. We conclude that both the measured $T_1$ and $T_2$ times as well as the long--term stability of the E' center at high density makes this defect an excellent candidate for local single spin measurements and in applications as probe spin system in spin--selection rule based spin readout schemes needed for silicon based spin quantum information or spintronics applications.


\begin{acknowledgements}
We acknowledge the support of this work by the National Science Foundation Major Research Instrumentation program (\#0959328 in support of K. A., C. W., and C. B), the Army Research Office (\#W911NF-10-1-0315 in support of A. P., C. W., and C. B.), and the National Science CAREER program (\#0953225 in support of D. P. W. and C. B.).
\end{acknowledgements}


\end{document}